\documentclass[english,aps,preprintnumbers,nofootinbib,superscriptaddress]{revtex4-1}
\usepackage{amssymb,amsmath}
\usepackage[usenames,dvipsnames]{xcolor}

\begin{document}
\preprint{YITP-16-105, IPMU16-0132}
\title{Where does curvaton reside?  Differences between bulk and brane frames}

\author{Fran\c{c}ois Larrouturou}
%\email{francois.larrouturou@ens.fr}
\affiliation{Ecole Normale Superieure, 45 rue d'Ulm, 75005 Paris, France}
\affiliation{Center for Gravitational Physics, Yukawa Institute for Theoretical Physics, Kyoto University, 606-8502, Kyoto, Japan}

\author{Shinji Mukohyama}
%\email{shinji.mukohyama@yukawa.kyoto-u.ac.jp}
\affiliation{Center for Gravitational Physics, Yukawa Institute for Theoretical Physics, Kyoto University, 606-8502, Kyoto, Japan}
\affiliation{Kavli Institute for the Physics and Mathematics of the Universe (WPI), The University of Tokyo Institutes for Advanced Study, The University of Tokyo, Kashiwa, Chiba 277-8583, Japan}

\author{Ryo Namba}
%\email{ryo.namba@ipmu.jp}
\affiliation{Kavli Institute for the Physics and Mathematics of the Universe (WPI), The University of Tokyo Institutes for Advanced Study, The University of Tokyo, Kashiwa, Chiba 277-8583, Japan}

\author{Yota Watanabe}
%\email{yota.watanabe@ipmu.jp}
\affiliation{Kavli Institute for the Physics and Mathematics of the Universe (WPI), The University of Tokyo Institutes for Advanced Study, The University of Tokyo, Kashiwa, Chiba 277-8583, Japan}
\affiliation{Center for Gravitational Physics, Yukawa Institute for Theoretical Physics, Kyoto University, 606-8502, Kyoto, Japan}

\date{\today}

\begin{abstract}

Some classes of inflationary models naturally introduce two distinct metrics/frames, and their equivalence in terms of observables has often been put in question. D-brane inflation proposes candidates for an inflaton embedded in the string theory and possesses descriptions on the brane and bulk metrics/frames, which are connected by a conformal/disformal transformation that depends on the inflaton and its derivatives. It has been shown that curvature perturbations generated by the inflaton are identical in both frames, meaning that observables such as the spectrum of cosmic microwave background (CMB) anisotropies are independent of whether matter fields---including those in the standard model of particle physics---minimally couple to the brane or the bulk metric/frame. This is true despite the fact that the observables are eventually measured by the matter fields and that the total action including the matter fields is different in the two cases. In contrast, in curvaton scenarios, the observables depend on the frame to which the curvaton minimally couples. Among all inflationary scenarios, we focus on two models motivated by the KKLMMT fine-tuning problem: a slow-roll inflation with an inflection-point potential and a model of a rapidly rolling inflaton that conformally couples to gravity. In the first model, the difference between the frames in which the curvaton resides is encoded in the spectral index of the curvature perturbations, depicting the nature of the frame transformation. In the second model, the curvaton on the brane induces a spectral index significantly different from that in the bulk and is even falsified by the observations. This work thus demonstrates that two frames connected by a conformal/disformal transformation lead to different physical observables such as CMB anisotropies in curvaton models.

\end{abstract}

\maketitle

\section{Introduction}

Inflation is one of the most powerful scenarios resolving the conceptual difficulties in the standard hot big bang cosmology and explaining the origin of the large-scale structure in the Universe~\cite{Guth:1980zm,Sato:1980yn,Starobinsky:1980te}, but it still lacks a convincing candidate for the inflaton, the field driving inflation. Among many models of inflation, those based on string theory have been attracting special attention, as they should in principle contain UV completion~\cite{Inflation_String_th}.

In higher-dimensional theories such as string theory, the position of an extended object in extra dimensions can be considered as a scalar field (or multiple of them) in the four-dimensional spacetime, provided that the object extends along the three uncompactified spatial dimensions of the four-dimensional spacetime. Brane inflation models take advantage of this simple fact. The first proposition to take the distance between two branes as a candidate for the inflaton was made by Dvali and Tye~\cite{Dvali:1998pa}. Those branes were then identified as a D-brane and anti-D-brane pair~\cite{Dvali:2001fw,Burgess:2001fx}. In the context of the brane inflation in type IIB string theory, one typically focuses on D3-branes and anti-D3-branes, but wrapped D5-branes can also be considered~\cite{Kobayashi:2007hm}.

In the string theory setup, however, it was found by Kachru-Kallosh-Linde-Maldacena-McAllister-Trivedi (KKLMMT) \cite{Kachru:2003sx} that slow-roll inflation is not as easy to achieve as previously thought, because of moduli stabilization. In spite of this difficulty, there are (at least) two ways to realize D-brane inflation in string theory: a slow-roll inflation driven by a potential with an inflection point~\cite{Baumann:2007np}, and a conformal rapid-roll inflation~\cite{Kofman:2007tr}. However, in the case of the inflection-point potential, a rather severe fine-tuning is required for the inflaton in the vicinity of the inflection point to produce cosmological perturbations consistent with observational data \cite{Baumann:2007np}. This situation may be ameliorated by a curvaton~\cite{Enqvist:2001zp,Lyth:2001nq,Moroi:2001ct}---another field responsible for the generation of curvature perturbations while being energetically subdominant during inflation---as the fine-tuning can be relaxed. In the case of conformal rapid-roll inflation, on the other hand, the introduction of a curvaton is compulsory since the rapidly rolling inflaton does not generate large-scale curvature perturbations \cite{Kofman:2007tr}. For these reasons, in both cases there is good motivation to consider the curvaton scenario in the context of D-brane inflation.

Working with branes in higher dimensions naturally introduces two distinct metrics/frames: one is the induced metric on a brane (or a stack of branes), and the other is the four-dimensional Einstein metric/frame. We shall call the former the {\it brane metric/frame} and the latter the {\it bulk metric/frame}. Physically speaking, the former is relevant to open string modes that propagate on the brane for which the induced metric is calculated. On the other hand, the bulk metric/frame is relevant to closed string modes that propagate in the bulk (as well as open string modes that propagate on other branes sitting at fixed positions in the extra dimensions).

The two metrics/frames are related to each other by a disformal transformation \cite{Bekenstein:1992pj} of the form $\tilde{g}_{\mu\nu}=A(\phi)g_{\mu\nu}+B(\phi)\partial_{\mu}\phi\partial_{\nu}\phi$, where $\tilde{g}_{\mu\nu}$ and $g_{\mu\nu}$ are the brane and bulk metrics, respectively, and $\phi$ is the inflaton.%
\footnote{Disformal transformations are also made use of in contexts other than string cosmology, e.g., cosmology with scalar-tensor theories or infrared modified theories of gravity. See Refs.~\cite{Deruelle:2010ht,Gong:2011qe,Chiba:2013mha,Motohashi:2015pra,Domenech:2015hka} and references therein.}
It has been known that curvature perturbations generated by the inflaton are invariant under disformal transformations of this type~\cite{Deruelle:2010ht,Gong:2011qe,Chiba:2013mha,Motohashi:2015pra,Domenech:2015hka},\footnote{See Ref.~\cite{Domenech:2015tca} for the invariance of the number of physical degrees of freedom under disformal transformations when $A$ and $B$ also depend on $g^{\mu\nu}\partial_\mu\phi\partial_\nu\phi$.} meaning that observables such as the spectrum of cosmic microwave background anisotropies are independent of whether matter fields---including those in the standard model of particle physics---minimally couple to the brane or to the bulk metric/frame.
This is nontrivially the case despite the fact that the observables are eventually measured by the matter fields and that the total action including the matter fields is different in the two cases. On the other hand, if a second field such as a curvaton is introduced, then observables may depend on whether it is (minimally) coupled to the bulk or the brane metric. Hence it can in principle act as a physical measure to differentiate the two different frames.

The aim of the present paper is therefore to investigate if inflationary observables can in principle differentiate the effects of a curvaton that is introduced in the bulk frame from those in the brane frame. 
We show how the scalar spectral index differs between the two models. As for the inflationary background, we take two concrete examples---motivated by the D-brane Dirac-Born-Infeld (DBI) cosmology---for each of the two curvaton models: slow-roll D-brane inflation driven by a potential with an inflection point~\cite{Baumann:2007np}, and conformal rapid-roll inflation~\cite{Kofman:2007tr}. Hence we calculate the curvature power spectra for both the bulk and brane curvatons in each of these inflationary models and explore the frame dependence.

The rest of the paper is organized as follows. In Sec.~\ref{sec:model} we first explain basic ingredients of the models, i.e., brane inflation models, disformal transformations, and the curvaton scenario. By combining them, we then describe the models that are studied in the present paper. In Sec.~\ref{sec:inflection}, we study slow-roll inflation driven by a potential with an inflection point, introducing a curvaton in each frame. In Sec.~\ref{sec:rapidroll}, we analyze each curvaton model in conformal rapid-roll inflation. Section \ref{sec:summary} is then devoted to a summary of this paper and some discussions.

\section{Model description}
\label{sec:model}

\subsection{D-brane inflation}\label{sec_model_DBI}

D-brane inflation arises when a stack of $N$ D3-branes moves towards the tip of a warped throat (such as the Klebanov-Strassler (KS) throat \cite{Klebanov:2000hb} or any other geometries that are approximated by $AdS_5 \times X_5$, where $X_5$ is a five-dimensional Einstein manifold~\footnote{Away from the tip, the KS throat is well approached by a $AdS_5 \times S^5$ geometry.}), while some anti-D-branes are stacked at the tip where they minimize their energy~\cite{Inflation_String_th}. The present paper focuses only on D3-brane inflation, but it is straightforward to extend the analysis to D5- and D7-brane DBI inflation (see, e.g., Ref.~\cite{Kobayashi:2007hm}). In the throat region, the line element is given by 
\begin{equation}
 \mathrm{d} s_{10}^{2} \equiv g_{\alpha \beta}^{(10)} \mathrm{d} X^{\alpha} \mathrm{d} X^{\beta}
= h^{2}(y)\eta_{\mu \nu} \mathrm{d} x^{\mu} \mathrm{d} x^{\nu} \: +
\: h^{-2}(y) \left( \mathrm{d} y^{2} + y^{2} \mathrm{d} \Omega_{X_5}^2 \right),
\label{eq_model_ds_10_flat}
\end{equation}
where $\eta_{\mu \nu}=\mathrm{diag}(-1,1,1,1)$ is the four-dimensional Minkowski metric, $y$ is the radial coordinate in the extra dimensions, $\mathrm{d} \Omega_{X_5}^2$ is a $y$-independent metric on $X_5$, and $h(y)$ is a dimensionless positive function of $y$ called a warp factor. In order to avoid a curvature singularity, the warp factor should remain positive all the way down to the tip of the throat at $y=0$. On the other hand, for large enough $y$, the warp factor behaves as $h(y) \simeq y/L$ with $L^4 = 4 \pi^4 g_s N \alpha'^2/\mathrm{Vol}(X_5)$, where $g_s$ is the string coupling and $\alpha'$ is the Regge slope, so that the geometry is well approximated by $AdS_5 \times X_5$. In the present paper, we shall only consider the region far from the tip, and thus $h(y) \simeq y/L$ is a good approximation. We nonetheless introduce a tiny deviation from this behavior and parametrize it by a small parameter $\beta$ as $h(y)\simeq (y/L)^{1-\beta}$ (see Eq.~(\ref{eq_model_h_beta})). While the four-dimensional metric in supergravity solutions such as the KS throat \cite{Klebanov:2000hb} is set to be Minkowski, it can be promoted to a four-dimensional curved metric, provided that the extra dimensions are compactified and that all moduli are stabilized to a scale sufficiently higher than the physical scale of interest. We thus replace $\eta_{\mu\nu}$ in Eq.~(\ref{eq_model_ds_10_flat}) by a four-dimensional curved metric $g_{\mu \nu}$,
\begin{equation}
 \mathrm{d} s_{10}^{2} \equiv g_{\alpha \beta}^{(10)} \mathrm{d} X^{\alpha} \mathrm{d} X^{\beta}
= h^{2}(y)g_{\mu \nu} \mathrm{d} x^{\mu} \mathrm{d} x^{\nu} \: +
\: h^{-2}(y) \left( \mathrm{d} y^{2} + y^{2} \mathrm{d} \Omega_{X_5}^2 \right) \; ,
\label{eq_model_ds_10}
\end{equation}
with the $X_5$ part unchanged.

We now consider a D3-brane in the geometry (\ref{eq_model_ds_10}). We suppose that the D3-brane extends to the four-dimensions $\{x^{\mu}\}$ and that the angular coordinates of $X_5$ stay constant on the world-volume. Introducing the embedding function $Y(x)$ to describe the brane world volume as $y=Y(x)$, the induced metric on the world volume is then given by 
\begin{equation}
 \tilde{g}_{\mu \nu} = h^{2}(Y)g_{\mu \nu}  + h^{-2}(Y)\partial_\mu Y \, \partial_\nu Y
  = h^{2}(\phi)g_{\mu \nu}  + T_3^{-1}h^{-2}(\phi)\partial_\mu \phi \, \partial_\nu \phi,
  \label{eq_model_gmnt}
\end{equation}
where we have defined the inflaton field as $\phi = \sqrt{T_3}Y$, $T_3 \equiv 1/(2\pi)^3g_s \alpha'^2$ is the D3-brane tension, and we have denoted $h(Y)=h(\phi/\sqrt{T_3})$ as $h(\phi)$ for brevity. Assuming that the dilaton is stabilized to a constant value, the DBI action is written in terms of this induced metric as
\begin{equation}
S_{DBI}
\equiv -T_3 \int \! \mathrm{d}^{4} x \sqrt{-\mathrm{det}(\tilde{g}_{\mu \nu})}
= - T_3 \int \! \mathrm{d}^{4} x \sqrt{-g} h^{4}(\phi) \sqrt{1 + \frac{g^{\mu\nu}\partial_{\mu} \phi \, \partial_{\nu} \phi}{T_{3}h^{4}(\phi)} },
\label{eq_model_S_DBI}
\end{equation}
where $g=\mathrm{det}(g_{\mu \nu})$. Adding a Chern-Simons term, the inflaton potential, and the four-dimensional Einstein-Hilbert action, the four-dimensional low-energy effective action for the inflaton is 
\begin{equation}
S = 
\int \! \mathrm{d} ^{4} x \sqrt{-g}
\left[ \frac{1}{2} M_\mathrm{Pl}^{2} \mathcal{R} - T_{3} h^{4}(\phi) \sqrt{1 + \frac{g^{\mu\nu}\partial_{\mu} \phi \, \partial_{\nu} \phi}{T_{3}h^{4}(\phi)} }
+ T_{3} h^{4}(\phi) - V(\phi) \right],
\label{eqn:inflaton_action}
\end{equation}
where $M_\mathrm{Pl}$ is the reduced Planck mass and $\mathcal{R}$ is the Ricci scalar. 
Here we have separated the factor $T_3 h^4$ from the inflaton potential so that the action reduces to $\int d^4x \sqrt{-g} \left[ M_{\rm Pl}^2 {\cal R} /2 - V(\phi) \right]$ for a constant $\phi$.
For large enough $\phi$, the warp factor becomes $h(\phi)\simeq\phi/l$ with $l^4 = T_3^2 L^4 = \pi N T_3 / [2 \mathrm{Vol}(X_5)]$ so that the geometry is well approximated by $AdS_5 \times X_5$. In the present paper we shall consider a deviation from the $AdS_5$ warp factor, parametrized by a small parameter $\beta$ as
\begin{equation}
h(\phi) = \left( \frac{\phi}{l} \right)^{1-\beta}. 
\label{eq_model_h_beta}
\end{equation}
The $AdS_5 \times X_5$ geometry is recovered in the limit $\beta \rightarrow 0$.

\subsubsection{Slow-roll inflation}

When the derivative of the inflaton is small enough so that  $\left| g^{\mu\nu}\partial_{\mu}\phi\partial_{\nu}\phi\right| \ll T_3 h^4(\phi)$, the action (\ref{eqn:inflaton_action}) is reduced to that for a minimally coupled canonical scalar field as
\begin{equation}
S = 
\int \! \mathrm{d} ^{4} x \sqrt{-g}
\left[ \frac{1}{2} M_\mathrm{Pl}^{2} \mathcal{R} -\frac{1}{2}g^{\mu\nu}\partial_{\mu} \phi \, \partial_{\nu} \phi - V(\phi) \right].
\end{equation}
For a flat Friedmann-Lema\^{i}tre-Robertson-Walker (FLRW) background metric $g_{\mu\nu} = {\rm diag} \left( -1 , a^2 , a^2 , a^2 \right)$, where the scalar factor $a(t)$ is a function only of time $t$, the Friedmann equation and the equation of motion (EOM) of $\phi$ read, respectively,
\begin{eqnarray}
&& H^2 = \frac{1}{3 M_{\rm Pl}} \left[ \frac{1}{2} \left( \partial_t \phi \right)^2 + V(\phi) \right] \; ,
\label{Friedmann}\\
&& \partial_t^2 \phi + 3H \partial_t \phi + V_\phi = 0 \; ,
\label{eom-phi}
\end{eqnarray}
where $H$ is the Hubble parameter $H \equiv \partial_t a / a$, and $\phi$ in the subscript denotes a derivative with respect to $\phi$. When $\phi$ rolls down its potential sufficiently slowly, i.e., if the so-called slow-roll parameters defined with respect to $V(\phi)$ as
\begin{equation}
\epsilon \equiv \frac{M_{\rm Pl}^2}{2} \left(\frac{V_\phi}{V}\right)^2 \; ,\quad
\eta_\phi \equiv M_{\rm Pl}^2 \frac{V_{\phi\phi}}{V} \; ,
\label{SR-phi}
\end{equation}
are sufficiently small, then the motion of $\phi$ drives a (quasi--)de Sitter expansion, i.e. inflation, and Eqs.~\eqref{Friedmann} and \eqref{eom-phi} approximately become
\begin{equation}
H^2 \simeq \frac{V(\phi)}{3M_{\rm Pl}^2} \; , \quad
\partial_t\phi \simeq - \frac{V_\phi}{3 H} \; .
\label{Friedmann_app}
\end{equation}
In the slow-roll approximation, 
$\epsilon$ and $\eta_\phi$ in Eq.~\eqref{SR-phi} coincide at the leading order with the ``slow-roll'' parameters in terms of $H$, defined as
\begin{equation}
\epsilon_H \equiv - \frac{\partial_t H}{H^2} \; , \quad 
\eta_\phi^H \equiv \frac{\vert V_{\phi \phi}\vert}{3H^2} \; .
\label{SR-H}
\end{equation}
Thus we will hereafter use the two definitions (\ref{SR-phi}) and (\ref{SR-H}) interchangeably as long as the slow-roll approximations are valid and the leading-order contributions are involved.

\subsubsection{KKLMMT fine-tuning problem}

A naive expectation for the form of the potential $V(\phi)$ would be the Coulomb potential of interaction between a D3-brane and an anti-D3-brane \cite{Dvali:1998pa}:
\begin{equation}
V_C(\phi) = V_0\left[1-\left(\frac{M_{\rm Pl}\Delta}{\phi}\right)^4 \right],
\label{eq_model_V}
\end{equation}
where $\Delta$ is a dimensionless constant depending on $X_5$, $V_0 = 2 T_3 h_t^4$, and $h_t$ is the value of the warp factor at the tip of the throat. The constant $V_0$ term arises through the charge carried by the branes, and the $\phi$-dependent term is the interaction between them. This potential could be useful for inflation for two reasons: its flatness, which allows slow-roll inflation for sufficiently large but sub-Planckian values of $\phi$; and its positivity, which drives inflation (as $H^2 \propto V$). However, a problem arises due to moduli stabilization, known as the KKLMMT fine-tuning problem~\cite{Kachru:2003sx}: a D3-brane backreacts on the K\"{a}hler potential through its position, and hence the K\"{a}hler moduli is no longer the compactification volume itself but becomes a combination of the compactification volume and the inflaton. Upon stabilizing the K\"{a}hler moduli, this yields a $\phi$-dependent compactification volume, which induces a nonperturbative correction to $V(\phi)$,
\begin{equation}
V(\phi) = \frac{V_C(\phi)}{(1 - \phi^2/3M_{\rm Pl}^2)^2} \simeq V_C(\phi) + \frac{2}{3}\frac{V_C(\phi)}{M_{\rm Pl}^2} \phi^2, \label{eqn:KKLMMTpotential}
\end{equation}
and thus an effective mass squared for large $\phi$ is $m_\mathrm{eff}^2 \simeq (2/3)(V_0/M_{\rm Pl}^2) \simeq 2 H^2$, which spoils slow-roll inflation as $\eta_\phi \simeq 2/3$.

\subsubsection{Inflection point potential}
\label{subsubsec:inflection_model}

A way to avoid this large mass is to consider various other contributions to the potential and to fine-tune them against the contribution in Eq.~(\ref{eqn:KKLMMTpotential}). In a UV-complete setup such as string theory this is not as trivial as one might naively think. It may have been the case that all other contributions may also give positive contributions to the inflaton mass squared. It is thus important to show that a concrete string theory setup can provide negative corrections (and perhaps positive corrections as well) to the inflaton mass squared, making it possible to cancel the positive mass squared in Eq.~(\ref{eqn:KKLMMTpotential}). For example, one can consider a set of $n > 1$ D7-branes supersymmetrically wrapping a four-cycle that extend to the throat~\cite{Baumann:2007np}. For a Kuperstein embedding~\cite{Kuperstein:2004hy} of those branes, the new contribution to the effective single-field potential of the inflaton has a local minimum away from the tip. By adding this to Eq.~(\ref{eqn:KKLMMTpotential}) and fine-tuning the position of the D7-branes, the potential can develop an {\it inflection point} at some $\phi = \phi_0$, so that the effective potential can be expanded as
\begin{equation}
 V = V_0 + \lambda_1 (\phi - \phi_0) + \frac{1}{3!} \lambda_3 (\phi - \phi_0)^3 + \cdots \, , \quad
  \lambda_1 \geq 0,
\label{eq_model_inflection_potential}
\end{equation}
where dots represent terms of higher order in $(\phi-\phi_0)$. Hereafter, we assume that $\lambda_1\geq 0$ so that the brane near the inflection point tends to move towards the tip of the throat. In this case, $\eta_\phi$ vanishes near the inflection point, i.e., the large mass is canceled by the effect of the D7-branes, and one recovers slow-roll inflation near $\phi_0$, as long as $\epsilon(\phi_0)$ is small. In this case, the EOMs in the vicinity of $\phi_0$ recover Eq.~\eqref{Friedmann_app}.

This model requires a rather severe fine-tuning, in order to ensure a scalar spectral index that is consistent with the observational constraint $n_s =0.968 \pm 0.006$ \cite{Ade:2015lrj}. The fine-tuning is so severe that, once the potential is tuned, the total number of $e$-folds is inevitably more than $150$~\cite{Inflation_String_th}. For a small-field inflation scenario, this is a rather large number. While there is no observational upper bound on the total number of $e$-foldings, this large number at least illustrates that the fine-tuning is not easy to achieve. Indeed, if for example the total number of $e$-foldings is just enough to solve the homogeneity and flatness problems and if the inflaton is the leading contribution to primordial curvature perturbations, then the scalar spectral index is far from the observational bound. On the other hand, if we implement some additional mechanism (such as curvaton scenario) to generate primordial curvature perturbations, then the amount of fine-tuning may be significantly relaxed.

\subsubsection{Rapid-roll inflaton}
\label{subsubsec:rapidroll_model}

Another way to solve the problem of the large mass is to consider the large mass as a result of a conformal coupling of the inflaton to the curvature~\cite{Kofman:2007tr}. When a field couples to gravity nonminimally through the term of the form $-\xi\mathcal{R}\phi^2/2$, the field acquires an effective mass squared $\xi \mathcal{R}$. The case with $\xi = 1/6$ is called a conformal coupling. In a FLRW background, this effective mass squared $6\xi(\partial_t H + 2H^2)$ is of the order of $2H^2$ for the conformal coupling and this recovers the large mass arising from the moduli stabilization~\footnote{See, e.g., the Appendix of Ref.~\cite{Kachru:2003sx}.}. We thus consider an inflaton $\phi$ described by the action 
\begin{equation}
S_{rr} = \frac{1}{2} \int \, \mathrm{d} ^{4} x \sqrt{-g} \left[ M_\mathrm{Pl}^{2} \mathcal{R} - \partial^{\mu} \phi \, \partial_{\mu} \phi
-  2V(\phi) - \frac{1}{6} \mathcal{R} \phi^2 \right], 
\label{eq_model_rr_action}
\end{equation}
where the potential $V(\phi)$ is given by the Coulomb potential (\ref{eq_model_V}), i.e., $V(\phi)=V_C(\phi)$. The KKLMMT correction to the potential has already been taken into account by the conformal coupling and thus should not be added to the Coulomb potential (\ref{eq_model_V}).

An interesting feature of this model is that the Hubble expansion rate remains almost constant, leading to an almost de Sitter expansion, as long as the potential $V(\phi)$ is flat enough. Namely, the large mass due to the conformal coupling (and thus the KKLMMT correction due to the moduli stabilization) does not spoil the quasi--de Sitter expansion. The flatness of the potential is measured by the standard slow-roll parameter $\epsilon$ in Eq.~\eqref{SR-phi} and two additional parameters $\bar{\epsilon}$ and $\eta_c$ defined as~\footnote{The quantity $\bar{\epsilon}$ defined here corresponds to $\tilde{\epsilon}$ defined in Ref.~\cite{Kofman:2007tr}. In the present paper we shall use the notation $\tilde{\epsilon}$ to define something else. See Eq.~(\ref{eq_model_sr_t}).}
\begin{equation}
 \bar{\epsilon} \equiv \frac{V_{\phi}\phi}{2V}, \quad
  \eta_c \equiv \eta_{\phi} + \frac{c+2}{3}\left[\frac{V_{\phi\phi}\phi}{V_{\phi}}+c\right],
  \label{eqn:barepsilon-etac}
\end{equation}
where $\epsilon$ and $\eta_{\phi}$ are defined by Eq.~(\ref{SR-phi}), and $c$ ($\ne -2$) is a constant chosen so that $|\eta_c|$ is minimized in a range of $\phi$. For example, if $V_{\phi\phi}\phi/V_{\phi}$ stays approximately constant and if $|\eta_{\phi}|$ is small, then one can choose $c=-V_{\phi\phi}\phi/V_{\phi}$. Under the conditions
\begin{equation}
 \epsilon \ll 1, \quad |\bar{\epsilon}|\ll 1, \quad |\eta_c| \ll 1,
\end{equation}
the Friedmann equation and the EOM of $\phi$ for the flat FLRW background with a homogeneous $\phi$ become
\begin{equation}
H^2 \simeq \frac{V}{3M_{\rm Pl}^2},
\quad
\partial_t \phi + H\phi \simeq \frac{-V_\phi}{(2+c)H}.
\label{eq_model_rr_EOM}
\end{equation}
The solution at the zeroth order in ($\epsilon$, $\bar{\epsilon}$, $\eta_c$) is thus
\begin{equation}
 H^2 \simeq \frac{V_0}{3M_{\rm Pl}^2} \simeq \mathrm{const} ,
  \quad
\phi \simeq \phi_0 \frac{a_0}{a}.
\label{eq_model_rr_phi}
\end{equation}
Unlike slow-roll inflation, $\partial_t\phi$ is not small: this model is therefore called {\it conformal rapid-roll inflation}~\cite{Kofman:2007tr} and is realized if ($\epsilon$, $\bar{\epsilon}$, $\eta_c$) are simultaneously small~\footnote{In principle, the standard slow-roll parameter $\eta_{\phi}$ does not have to be small as long as ($\epsilon$, $\bar{\epsilon}$, $\eta_c$) are small. In the concrete example studied in the present paper, however, $\eta_{\phi}$ is also small.}.

The conformal rapid-roll inflation takes into account the KKLMMT correction to the potential due to the moduli stabilization and yet provides inflation without further fine-tuning. The model is a low-scale inflation and predicts a small number of $e$-foldings: $N_* \lesssim 34$~\cite{Kofman:2007tr} . However, this model itself does not generate cosmological curvature perturbations and thus has to be completed with another generation mechanism, such as the curvaton scenario.

\subsection{Disformal transformation}
\label{subsec:disformal-tr}

Disformal transformation was first introduced in Ref.~\cite{Bekenstein:1992pj} as a relation between two metrics that respects the weak equivalence principle and causality. In the context of D-brane models in our current study, the transformation of the form \eqref{eq_model_gmnt} gives a relation between the induced metric on the brane $\tilde{g}_{\mu\nu}$ and the four-dimensional Einstein metric $g_{\mu\nu}$. We call the former the {\it brane metric/frame} and the latter the {\it bulk metric/frame}. As we shall see now, the brane metric/frame is relevant to open string modes that propagate on the brane for which the induced metric is calculated. On the other hand, the bulk metric/frame is relevant to closed string modes that propagate in the bulk as well as open string modes that propagate on other branes sitting at fixed positions in the extra dimensions.

In order to see that open string modes on the D3-brane propagate freely on the brane metric/frame, let us consider the $U(1)$ gauge field on the brane. The $U(1)$ gauge field is described by the DBI action, 
\begin{equation}
 I_{\mathrm{DBI}} = -T_3\int d^4x
  \sqrt{-\det(\tilde{g}_{\mu\nu}+2\pi\alpha'F_{\mu\nu})} \; ,
\end{equation}
where $F_{\mu\nu}$ is the field strength of the $U(1)$ gauge field on the brane. Here, for simplicity we have assumed that the pullback of the Neveu-Schwarz(NS)---NS antisymmetric field on the brane world volume vanishes and that the dilaton is stabilized as in the construction by Giddings, Kachru, and Polchinski~\cite{Giddings:2001yu}. Expanding the DBI action with respect to $F_{\mu\nu}$, it is easy to see that the $U(1)$ gauge field on the brane minimally couples to the brane metric $\tilde{g}_{\mu\nu}$ (instead of $g_{\mu\nu}$). Also, by splitting $\phi$ into the background and perturbation as $\phi=\phi^{(0)}+\delta\phi$, one can easily see that the perturbation $\delta\phi$ couples minimally to the background brane metric $\tilde{g}^{(0)}_{\mu\nu}=h^{2}(\phi^{(0)})g_{\mu \nu}+T_3^{-1}h^{-2}(\phi^{(0)})\partial_\mu \phi^{(0)} \, \partial_\nu \phi^{(0)}$. Since the $U(1)$ gauge field and the brane bending mode $\delta\phi$ are open string degrees of freedom, we can thus regard $\tilde{g}_{\mu\nu}$ as the metric relevant to the open string modes propagating on the brane.

On the other hand, the bulk metric/frame $g_{\mu\nu}$ is relevant to closed string modes that propagate in the bulk as well as open string modes that propagate on other branes sitting at fixed positions in the extra dimensions. Indeed, assuming that all moduli are properly stabilized, four-dimensional low-energy modes of these degrees of freedom propagate on $g_{\mu\nu}$. Also, after compactification of extra dimensions and stabilization of all moduli, the four-dimensional Einstein-Hilbert action for $g_{\mu\nu}$ is obtained from the ten-dimensional Einstein-Hilbert action. This is of course the reason why we called $g_{\mu\nu}$ the four-dimensional Einstein metric.

Despite the distinct physical meanings of the two metrics/frames explained above, the disformal transformation leaves some inflationary observables invariant if we consider $\phi$ as the inflaton and if it is the dominant source of primordial curvature perturbations~\cite{Deruelle:2010ht,Gong:2011qe,Chiba:2013mha,Motohashi:2015pra,Domenech:2015hka}. So, if the primordial curvature perturbations are generated by the inflaton, they are not affected by the disformal transformation.

However, if primordial curvature perturbations arise due to another field (such as a curvaton field) $\sigma$, then the transformation acquires a physical meaning: the way $\sigma$ interacts with the inflaton $\phi$ depends crucially on whether $\sigma$ is introduced in the bulk metric/frame or in the brane metric/frame. By ``being introduced in the bulk metric/frame'', we mean that the Lagrangian density of $\sigma$ is $\mathcal{L}^\sigma = \sqrt{-g}[- g^{\mu \nu} \partial_{\mu} \sigma \, \partial_{\nu} \sigma/2 - U(\sigma)]$, assuming  that $\sigma$ is a canonical scalar. On the other hand, by ``being introduced in the brane metric/frame'' we mean that the Lagrangian density of $\sigma$ is $\tilde{\mathcal{L}}^\sigma = \sqrt{-\tilde{g}}[- \tilde{g}^{\mu \nu} \partial_{\mu} \sigma \, \partial_{\nu} \sigma/2 - U(\sigma)]$. In this case the spectrum of primordial curvature perturbations depends on the frame in which $\sigma$ is introduced. 
An explicit demonstration of this is the aim of the present paper.

For a FLRW background with a homogeneous inflaton,
\begin{equation}
 g_{\mu\nu}\mathrm{d} x^\mu \mathrm{d} x^\nu=-\mathrm{d} t^2+a(t)^2\delta_{ij}\mathrm{d} x^i \mathrm{d} x^j, \quad
  \phi = \phi(t),
\end{equation}
the disformal transformation (\ref{eq_model_gmnt}) leads to
\begin{equation}
 \tilde{g}_{\mu\nu}\mathrm{d} x^\mu \mathrm{d} x^\nu=- \tilde{N}^2\mathrm{d} t^2+\tilde{a}^2\delta_{ij}\mathrm{d} x^i \mathrm{d} x^j, \quad
  \tilde{N} = \frac{h}{\gamma}, \quad
  \tilde{a} = ha,
  \label{eqn:tildegmunu_FLRW}
\end{equation}
where we have defined the Lorentz-like factor $\gamma$~\footnote{In term of $\phi$, the line element (\ref{eq_model_ds_10}) yields a speed of light along extra dimensions $c_{6} = \sqrt{T_3}h^2$.} as
\begin{equation}
 \gamma \equiv \left[1 + \frac{(\partial \phi)^2}{T_3h^4}\right]^{-1/2}, \quad
  (\partial \phi)^2 \equiv g^{\mu \nu} \partial_\mu \phi \, \partial_\nu \phi = -(\partial_t\phi)^2, \label{eqn:def_gamma}
\end{equation}
and it is understood that $h=h(Y=\phi/\sqrt{T_3})$. The two functions $\tilde{N}$ and $\tilde{a}$ are the lapse function and the scale factor from the viewpoint of open string modes on the brane in the time coordinate $t$. One can then introduce the proper time $\tau$ on the brane as $\mathrm{d} \tau \equiv \tilde{N} \mathrm{d} t$ and the Hubble expansion rate on the brane $\tilde{H}$ as
\begin{equation}
 \tilde{H} \equiv \partial_{\tau} \ln (\tilde{a}) = \frac{\gamma}{h} H ( 1 - \epsilon_{h} ), \quad
  \epsilon_{h} \equiv -\frac{\partial_t h}{Hh}.
\label{eq_model_Ht}
\end{equation}
The quantity $\epsilon_{h}$ is positive, provided that the brane moves towards the tip of the throat.

\subsection{Curvaton scenario}
\label{subsec:curvaton}

When it is difficult or impossible for the inflaton to generate sufficient primordial curvature perturbations consistent with observational data (which could be the case in both of the presented models of inflation), one can introduce another field to produce them: the curvaton $\sigma$ \cite{Enqvist:2001zp,Lyth:2001nq,Moroi:2001ct}. The curvaton field is energetically subdominant during inflation, and its quantum fluctuations generate isocurvature perturbations (entropy perturbations without curvature perturbations) which will be later converted to curvature perturbations.

The curvaton scenario can be regarded as a special case of multifield inflation, and models of multifield brane inflation have been widely studied in the literature, e.g.~regarding the position of a D7-brane as the second field \cite{Baumann:2007ah}, and regarding six coordinates of the position of a D3-brane in the extra dimension as six fields \cite{Agarwal:2011wm, Dias:2012nf, McAllister:2012am}. The positions of other types of branes can also be other fields. The authors of Refs.~\cite{Baumann:2010sx, Gandhi:2011id} studied the effect of symmetries of the high-scale theory on coefficients of the effective theory of branes. Phenomenological consequences of multifield brane inflation were also studied in Refs.~\cite{Panda:2007ie, Krause:2007jk, Shandera:2006ax, Peiris:2007gz, Chen:2008ai}. The authors of Ref.~\cite{Chen:2008ada} treated multifield effects at the end of D-brane inflation. In those multifield inflationary models a second field affects the background geometry. On the contrary, the curvaton is a spectator field during inflation, meaning that it is energetically subdominant and does not affect the background geometry.  The curvaton scenario in brane inflation was considered in Ref.~\cite{Pilo:2004mg} where an axion-like field plays the role of curvaton. The authors of Refs.~\cite{Li:2008fma, Cai:2010rt} took the kinetic term of a curvaton as the DBI form.

In order to compute the observable quantities generated by the field as in Ref.~\cite{Lyth:2001nq}, for simplicity we assume that the curvaton is neither interacting nor washed out during inflation. We furthermore assume that there is a mass hierarchy $\eta_\sigma \equiv U_{\sigma \sigma}/(3H^2) \ll 1$, where $U$ is the potential of $\sigma$ and a subscript $\sigma$ represents a derivative with respect to $\sigma$, and that the curvaton energy density $\rho_\sigma$ becomes a non-negligible fraction of the total energy density some time after inflation and before the decay of the curvaton.

In the present paper we consider two distinct models of the curvaton scenario for each of the two presented models of inflation. In the first model, we introduce a curvaton field $\sigma$ in the bulk metric/frame $g_{\mu\nu}$ in the sense that was explained in Sec.~\ref{subsec:disformal-tr}. We call this a {\it bulk curvaton} model. The Lagrangian density is thus
\begin{equation}
 \mathcal{L}^\sigma = \sqrt{-g}\left[- \frac{1}{2}g^{\mu \nu} \partial_{\mu} \sigma \, \partial_{\nu} \sigma - U(\sigma)\right] \quad (\mbox{bulk curvaton model}),
\label{bulk_curv_action}
\end{equation}
where we have assumed that the curvaton $\sigma$ is a canonical scalar. In principle the action for the curvaton may involve nonlinear kinetic terms, such as the DBI kinetic term \cite{Li:2008fma} mimicking the kinetic part of Eq.~(\ref{eqn:inflaton_action}). However, in the curvaton scenario it is generically assumed that fluctuations of a curvaton correspond to isocurvature perturbations during inflation, meaning that $\dot{\sigma}^2 \ll \dot{\phi}^2$. As will be proven later, in both inflation scenarios considered in the present paper, the DBI correction to the standard canonical kinetic term for the inflaton, parametrized by $1-1/\gamma$, nearly vanishes. Therefore, unless the effective string scale for the curvaton field is significantly lower than that for the inflaton, the DBI correction should be negligibly small for the curvaton field. For this reason, in the context of the present paper one can safely adopt the canonical kinetic term as in the action (\ref{bulk_curv_action}). In the curvaton scenario it is also assumed in general that the curvaton is energetically subdominant and does not roll during inflation. For this reason we suppose that the curvaton action enjoys an approximate shift symmetry in the field regime relevant for the evolution during inflation. This in particular implies that the nonminimal coupling of the curvaton to the spacetime curvature is highly suppressed, i.e. the curvaton is almost (if not exactly) minimally coupled to the metric.

Examples of the bulk curvaton model are axions (pseudoscalar fields enjoying approximate shift symmetries) originated from ten-dimensional antisymmetric fields and any light fields living on other branes sitting either in the same throat or in other throats. See Ref.~\cite{Kobayashi:2009cm} for a concrete construction of the curvaton field from the angular position of another brane. The resulting curvature perturbations obey, assuming (approximately) constant $\eta_\sigma$,
\begin{equation}
{\cal P}_\zeta \simeq \left[ \frac{R}{6\pi} \, \frac{U_\sigma (\sigma_*)}{U(\sigma_*)} \, H_{*0} \right]^2 \left( \frac{k}{k_0} \right)^{n_s-1} \; , \quad
 n_s - 1 \simeq 2 \eta_{\sigma} - 2 \epsilon_H, \quad
 \eta_{\sigma} \equiv \frac{\vert U_{\sigma\sigma}\vert}{3H^2},
\label{Pzeta_gen}
\end{equation}
where $R = 3\rho_\sigma / (4 \rho_r + 3 \rho_\sigma)$ is the fraction of curvaton energy density at curvaton decay \cite{Lyth:2001nq}, $n_s$ is the spectral index, $k_0$ is a pivot scale, a subscript $*$ denotes horizon exit ($k = a_*H_*$), and a subscript $*0$ indicates that the quantity is to be evaluated when the pivot scale exits the horizon. For simplicity, we hereafter assume that the curvaton potential during inflation can be approximated by a quadratic one $U(\sigma) \simeq m_\sigma^2 \sigma^2 /2$, reducing Eq.~(\ref{Pzeta_gen}) to
\begin{equation}
\mathcal{P}_{\zeta} \simeq \left(\frac{R H_{*0}}{3 \pi \sigma_*} \right)^{2} \left( \frac{k}{k_0} \right)^{n_s-1}, \quad
 n_s - 1 \simeq 2 \eta_{\sigma} - 2 \epsilon_H, \quad
 \eta_{\sigma} \equiv \frac{m_\sigma^2}{3H^2}.
\label{eq_model_Pc}
\end{equation}

In the second model, on the other hand, we introduce a curvaton field $\sigma$ in the brane metric/frame $\tilde{g}_{\mu\nu}$ in the sense that was explained in Sec.~\ref{subsec:disformal-tr}. We call this a {\it brane curvaton} model. The Lagrangian density is thus
\begin{equation}
 \tilde{\mathcal{L}}^\sigma = \sqrt{-\tilde{g}}\left[- \frac{1}{2}\tilde{g}^{\mu \nu} \partial_{\mu} \sigma \, \partial_{\nu} \sigma - U(\sigma)\right], \quad (\mbox{brane curvaton model}).
\label{brane_curv_action}
\end{equation}
Here, for the reasons explained after Eq.~(\ref{bulk_curv_action}), we do not include the DBI correction to the standard canonical kinetic term nor the nonminimal curvature coupling of the curvaton. Examples of the brane curvaton model are the angular position of the inflaton brane~\cite{Kobayashi:2009cm} and any other light fields living on the inflaton brane (such as angles among intersecting branes). The brane metric $\tilde{g}_{\mu\nu}$ shows an inflationary behavior when
\begin{equation}
\tilde{\epsilon} \equiv - \frac{\partial_{\tau} \tilde{H}}{\tilde{H}^2} = \frac{1}{1 - \epsilon_{h}} \left( \epsilon + \epsilon_{\gamma} - \epsilon_{h} + \epsilon_{1 \! - \! \epsilon_h} \right)
\label{eq_model_sr_t}
\end{equation} 
is small, where $\epsilon_q \equiv - \partial_t q/(Hq)$ with $q = \left\{ \gamma , h , 1 - \epsilon_h \right\}$. This is realized when the parameters $\epsilon_q$ appearing in the expression for $\tilde{\epsilon}$ are all small or canceling with each other. Under this condition and if the mass of the curvaton is light enough, then the previous derivation applies similarly and---as the brane frame reduces to the bulk frame after inflation---one has (with the horizon exit at $k = \tilde{a}_* \tilde{H}_*$)
\begin{equation}
\tilde{\mathcal{P}}_{\zeta} \simeq \left(\frac{R \tilde{H}_{*0}}{3 \pi \sigma_*} \right)^{2} \left( \frac{k}{k_0} \right)^{\tilde n_s-1} \; ,
\quad 
\tilde{n}_s - 1 \simeq 2 \tilde{\eta}_{\sigma} - 2 \tilde{\epsilon},\quad
 \tilde{\eta}_{\sigma} \equiv \frac{m_\sigma^2}{3\tilde{H}^2}.
\label{eq_model_Pct}
\end{equation}

If the brane metric does not exhibit an inflationary behavior and/or if the mass of the curvaton is not light enough, then the formula (\ref{eq_model_Pct}) does not hold, and one has to reconsider the power spectrum based on the brane curvaton Lagrangian density,
\begin{equation}
 \tilde{\mathcal{L}}_\sigma = \sqrt{-\tilde{g}}\left[-\frac{1}{2}\tilde{g}^{\mu \nu} \partial_{\mu} \sigma \, \partial_{\nu} \sigma - U(\sigma)\right]
						= \frac{1}{2}\tilde{a}^2\left[(\partial_{\tilde{\eta}}\sigma)^2-\delta^{ij}\partial_i\sigma\partial_j\sigma\right] - \tilde{a}^4U(\sigma),
\end{equation}
and the corresponding EOM,
\begin{equation}
 \partial_{\tilde{\eta}}^2\sigma + 2(\partial_{\tilde{\eta}}\ln\tilde{a})\partial_{\tilde{\eta}}\sigma - \delta^{ij}\partial_i\sigma\partial_j\sigma + \tilde{a}^2V_{\sigma} = 0,
\label{eq_model_EOM_t}
\end{equation}
where $\tilde{\eta}$ is the brane conformal time defined by
\begin{equation}
\mathrm{d} \tilde{\eta} = \frac{\mathrm{d} \tau}{\tilde{a}} = \frac{\tilde{N}\mathrm{d} t}{\tilde{a}}
= \frac{dt}{a \gamma} \; ,
\label{def-etatil}
\end{equation}
using Eq.~\eqref{eqn:tildegmunu_FLRW}.

\subsection{Summary of model description}

Combining the basic ingredients reviewed in the previous subsections, we here describe the analysis we shall perform in the following sections. Our goal is to differentiate the effects of the curvaton in the bulk vs.~brane frames of a string-theoretical setup by comparing observable signatures, with a particular focus on the curvature power spectrum generated in the curvaton scenario. We investigate two distinct models of inflation motivated by the D-brane dynamics in string theory. We compute the curvature power spectra ${\cal P}_\zeta$ in the two frames and the difference in their spectral indices in each of these models. The models and the basic structure of our analysis are therefore as follows.
\begin{enumerate}
\item%[1.] 
{\it A slow-roll inflaton with an inflection-point potential}: Model description in Sec.~\ref{subsubsec:inflection_model}, and analysis in Sec.~\ref{sec:inflection}.

\begin{enumerate}
\item[(1-i)] ${\cal P}_\zeta$ by a curvaton in the bulk frame.
\item[(1-ii)] $\tilde{\cal P}_\zeta$ by a curvaton in the brane frame.
\end{enumerate}

\item[2.] {\it A rapid-roll inflaton conformally coupled to gravity}: Model description in Sec.~\ref{subsubsec:rapidroll_model}, and analysis in Sec.~\ref{sec:rapidroll}.

\begin{enumerate}
\item[(2-i)] ${\cal P}_\zeta$ by a curvaton in the bulk frame.
\item[(2-ii)] $\tilde{\cal P}_\zeta$ by a curvaton in the brane frame.
\end{enumerate}

\end{enumerate}

The two frames are related by the conformal/disformal transformation of the type \eqref{eq_model_gmnt} with the warp factor (\ref{eq_model_h_beta}), and the difference in the observables is featured by the relevant quantities of the transformation introduced in Secs.~\ref{subsec:disformal-tr} and \ref{subsec:curvaton}. The curvaton scenario generically assumes that the curvaton fluctuations correspond to isocurvature perturbations during inflation, meaning that the time derivative of the curvaton at the background level is much smaller than that of the inflaton. Thus, for the reason explained after Eq.~(\ref{bulk_curv_action}), we suppose that in each inflation model the curvaton $\sigma$ has a canonical kinetic term. For the reason that was also explained after Eq.~(\ref{bulk_curv_action}), it is supposed that the curvaton is almost (if not exactly) minimally coupled to each frame metric: see Eqs.~\eqref{bulk_curv_action} and \eqref{brane_curv_action}. We calculate the scalar spectral indices in all four cases and investigate observable differences due to the frames in which the curvaton $\sigma$ resides.

\section{Inflection-point models}
\label{sec:inflection}

\subsection{Range of parameters}

For the inflection-point potential (\ref{eq_model_inflection_potential}), we specify the range of parameters to be considered. First, the potential slow-roll parameters (\ref{SR-phi}) are calculated as
\begin{equation}
\epsilon \simeq \epsilon_0 \left( \frac{1 + \theta \varphi^{2}}{1 + \varphi + \frac{1}{3} \theta \varphi^{3} } \right)^{2},
\quad
\eta_{\phi} \simeq 4 \epsilon_0 \frac{\theta \varphi}{1 + \varphi + \frac{1}{3} \theta \varphi^{3}}, \label{eqn:epsilon-eta-inflection}
\end{equation}	
where
\begin{equation}
\epsilon_0 \equiv \frac{1}{2}\frac{M_{\rm Pl}^2 \lambda_1^2}{V_0^2}
\end{equation}
is the value of $\epsilon$ at the inflection point $\phi_0$ and we have introduced the dimensionless variable $\varphi$ and the dimensionless parameter $\theta$ as
\begin{equation}
 \varphi \equiv \frac{\lambda_1}{V_0}(\phi - \phi_0), \quad
  \theta \equiv \frac{V_0^2}{2}\frac{\lambda_3}{\lambda_1^3},
\end{equation}
so that the potential (\ref{eq_model_inflection_potential}) is rewritten as
\begin{equation}
 V(\varphi) = V_0 \left[1 + \varphi + \frac{1}{3}\theta\varphi^3 + O(\varphi^4)\right].
  \label{eqn:Vvarphi}
\end{equation} 
As the first condition, we thus demand that the slow-roll approximation be valid at the inflection point, i.e. $\epsilon_0\ll 1$, yielding
\begin{equation}
 0  \leq \lambda_1 \ll \frac{V_0}{M_{\rm Pl}}. 
\end{equation}

As the second condition, we also impose a microscopic constraint that applies to the string theory setup under consideration~\cite{Baumann:2006cd}. Assuming that the inflation starts at or before the inflection point, we require that the difference between the value of $\phi$ at the end of inflation and the value at the inflection point be smaller than the Planck scale in magnitude: 
\begin{equation}
 |\phi_{\rm end} - \phi_0| \lesssim M_{\rm Pl}.  \label{eqn:Deltaphi<MPl}
\end{equation}

The third condition is to ensure the validity of the truncated potential (\ref{eqn:Vvarphi}). Assuming that the potential (\ref{eqn:Vvarphi}) truncated at the third order in $\varphi$ is a good approximation to the genuine potential based on string theory all the way from the vicinity of the inflection point to the end of inflation, the end of inflation is specified by either $\epsilon=1$ or $\eta_{\phi}=\pm 1$, where $\epsilon$ and $\eta_{\phi}$ are given by Eq.~(\ref{eqn:epsilon-eta-inflection}). In the truncated potential (\ref{eqn:Vvarphi}) we now demand that the first term dominates over the sum of the second and third terms so that the truncation is justified. Since $\varphi_{\rm end}=\sqrt{2\epsilon_0}(\phi_{\rm end} - \phi_0)/M_{\rm Pl}$, the second term is always much smaller than the first term under the microscopic constraint (\ref{eqn:Deltaphi<MPl}) and the slow-roll condition, $\epsilon_0\ll 1$, that we have already imposed. In order for the first term to dominate the third term all the way down to the end of inflation, it is necessary and sufficient that either $|\theta\varphi_{\epsilon}^3/3|\ll 1$ or $|\theta\varphi_{\eta}^3/3|\ll 1$ is satisfied, where $\varphi_{\epsilon}$ and $\varphi_{\eta}$ are defined by $\epsilon_0(\theta\varphi_{\epsilon}^2)^2=1$ and $4\epsilon_0\theta\varphi_{\eta}=-1$, respectively. We thus obtain the condition $\epsilon_0^3\theta^2\gtrsim 1$, which yields
\begin{equation}
\lambda_3 \gtrsim \frac{V_0}{M_{\rm Pl}^3}.
\end{equation}
This implies that $|\varphi_{\eta}|<|\varphi_{\epsilon}|$ and that $\epsilon<|\eta_{\phi}|=O(1)$ at the end of inflation.

In summary, the regime of parameters to be considered is
\begin{equation}
0 \leq \lambda_1 \ll \frac{V_0}{M_{\rm Pl}},
\quad
\lambda_3 \gtrsim \frac{V_0}{M_{\rm Pl}^3},
\quad
\vert \phi_{\rm end} - \phi_0 \vert \lesssim M_{\rm Pl}.
\label{eq_inflection_bounds}
\end{equation}
The first and second conditions imply that $\theta\gg 1$. Slow-roll inflation with an inflection-point potential therefore requires a very flat potential with a strong inflection behavior, and a short range of $\phi$.

\subsection{Inflaton evolution}

In the regime of parameters (\ref{eq_inflection_bounds}), the Friedmann and continuity equations \eqref{Friedmann_app} during inflation read
\begin{equation}
H^2 \simeq \frac{V}{3 M_{\rm Pl}^2}
\simeq H_0^2,
\quad
\partial_t\varphi \simeq -\frac{V_{\phi}}{3 H}\frac{\lambda_1}{V_0}
\simeq - 2\epsilon_0 H_0 (1 + \theta \varphi^{2}), 
\end{equation}
where $H_0 \equiv \sqrt{V_0/(3 M_{\rm Pl}^2)}$. The second equation yields
\begin{equation}
\varphi(t) \simeq - \frac{1}{\sqrt{\theta}} \tan \left[2 \sqrt{\theta} \epsilon_0 H_0 (t \! - t_0) \right] \; ,
\end{equation}
where $t_0$ denotes the time when the inflaton passes through the inflection point of its potential.
The slow-roll parameters defined in Eq.~(\ref{SR-phi}) are
\begin{equation}
 \epsilon \simeq \frac{\epsilon_0}
  {\cos^4\left[2 \sqrt{\theta} \epsilon_0 H_0 (t \! - t_0) \right]}\,,
\quad 
\eta_\phi \simeq -4\epsilon_0\sqrt{\theta}\tan \left[2 \sqrt{\theta} \epsilon_0 H_0 (t \! - t_0) \right].
\label{eq_inflection_sr}
\end{equation}
Hence, although $|\eta_{\phi}|$ vanishes at the inflection point, it grows as the inflaton rolls. As we have seen in the previous subsection, it is $\eta_{\phi}$ that determines the end of inflation under the condition (\ref{eq_inflection_bounds}).

\subsection{Spectral indices in two models of the curvaton}

The scale factor $\tilde{a}$, the Hubble expansion rate $\tilde{H}$, and the first slow-roll parameter $\tilde{\epsilon}$ on the brane are given by Eqs.~(\ref{eqn:tildegmunu_FLRW}), (\ref{eq_model_Ht}), and (\ref{eq_model_sr_t}), respectively. We thus need to calculate $\epsilon_h$, $\epsilon_{\gamma}$, and $\epsilon_{1 \! - \! \epsilon_h}$. By using $\partial_t\phi/H =  - M_{\rm Pl} \sqrt{2\epsilon_H} \simeq - M_{\rm Pl} \sqrt{2 \epsilon}$, $\partial_t{\epsilon}/(H\epsilon) \simeq 4 \epsilon - 2 \eta_\phi$ and the form of the warp factor (\ref{eq_model_h_beta}), it is easy to obtain 
\begin{equation}
\epsilon_{h} 
\simeq (1 -\beta)\frac{M_{\rm Pl}}{\phi}\sqrt{2\epsilon}, 
\quad
\epsilon_{\gamma}
\simeq (\gamma^2-1)(\eta_\phi -\epsilon - 2 \epsilon_{h}),
\quad
\epsilon_{1 - \epsilon_h} 
\simeq \frac{\epsilon_{h}}{1-\epsilon_{h}}\left(\frac{\epsilon_{h}}{1 \! - \! \beta} + 2\epsilon - \eta_\phi \right).
\end{equation}
Considering far-from-tip inflation (i.e., not too small $\phi/M_{\rm Pl}$) and only small (if any) deviations from the $AdS_5$ throat (i.e., small $\beta$), we see that $\epsilon_h\ll 1$ and that $\epsilon_{1 \! - \! \epsilon_h}$ is negligible compared with $\epsilon_{h}$. Moreover, since we are considering slow-roll inflation, $\gamma = 1 + \mathcal{O}(\epsilon)$ and thus $\epsilon_{\gamma}$ is second-order in slow-roll parameters. We thus have
\begin{equation}
 \tilde{a}  = ha, \quad
 \tilde{H} \simeq  \frac{H}{h},\quad
 \tilde{\epsilon} \simeq \epsilon - (1-\beta )\frac{M_{\rm Pl}}{\phi}\sqrt{2\epsilon}.
 \label{eqn:tildea-tildeH^tildeepsilon}
\end{equation}
A quasi--de Sitter expansion, i.e., $|\tilde{\epsilon}|\ll 1$, is therefore realized in the brane frame $\tilde{g}_{\mu \nu}$ if $\phi_0/M_{\rm Pl}$ is not too small. Since this work deals with far-from-tip inflation, it is reasonable to assume that this is the case.

In the brane curvaton model, a mode with a comoving wave number $k$ exits the horizon when $k=\tilde{a}\tilde{H}$. On the other hand, in the bulk curvaton model a mode with the same comoving wave number $k$ exits the horizon when $k=aH$. Therefore, in principle the time of horizon exit may be different in the two models of the curvaton. However, we can easily show that in practice the difference is negligible. By definition we have $\tilde{a}_{\tilde{*}}\tilde{H}_{\tilde{*}}=a_*H_*$, where the subscripts $\tilde{*}$ and $*$ represent the horizon exit of the mode with comoving momentum $k$ in the brane curvaton model and in the bulk curvaton model, respectively. By using Eq.~(\ref{eqn:tildea-tildeH^tildeepsilon}), this is reduced to $a_{\tilde{*}}H_{\tilde{*}}\simeq a_*H_*$, leading to
\begin{equation}	
 t_{\tilde{*}} \simeq t_*.
\end{equation}
Thus in the two models of the curvaton, modes with the same comoving wave number exit the horizon at almost the same time.

Since a quasi--de Sitter expansion is realized in both frames, the spectral indices are given by Eqs.~(\ref{eq_model_Pc}) and (\ref{eq_model_Pct}). The difference between them, $\tilde{n}_s-n_s$, is calculated as 
\begin{equation}
 \tilde{n}_s - n_s \simeq 2\left(\frac{H_*^2}{\tilde{H}_*^2}-1\right)\eta_{\sigma} - 2(\tilde{\epsilon}-\epsilon)
  \simeq \left[\left(\frac{\phi_*}{l}\right)^{2(1-\beta)}-1\right]\eta_{\sigma} +  2(1-\beta )\frac{M_{\rm Pl}}{\phi_*}\sqrt{2\epsilon},
\end{equation}
where we have used Eq.~(\ref{eqn:tildea-tildeH^tildeepsilon}) and the form of the warp factor (\ref{eq_model_h_beta}). For the $AdS_5$ throat ($\beta=0$), we obtain
\begin{equation}
 \tilde{n}_s - n_s \simeq \left[\left(\frac{\phi_*}{l}\right)^2-1\right]\eta_{\sigma} +  \frac{2M_{\rm Pl}}{\phi_*}\sqrt{2\epsilon}. \label{eqn:ns-nstilde-inflection-AdS5}
\end{equation}
Therefore in the scenario under consideration, the frame transformation has a physical meaning in the sense that the brane curvaton model and the bulk curvaton model predict different spectral indices even if the curvaton mass is the same in the two models. In other words, if the curvaton is responsible for the primordial curvature perturbations as observed in the cosmic microwave background radiation, then the inflaton potential, the curvaton mass, and the warp factor are constrained by observation in different ways, depending on whether the curvaton is introduced in the brane frame or in the bulk frame. In either case the introduction of a curvaton significantly relaxes the fine-tuning of the inflaton potential. For example, the original model without a curvaton is inconsistent with observation if the total number of $e$-foldings is just enough to solve the homogeneity and flatness problems \cite{Baumann:2007np}. On the other hand, this difficulty does not apply to the model with a curvaton.

\section{Conformal rapid-roll models}
\label{sec:rapidroll}

\subsection{Domain of rapid-roll inflation}\label{sec_rr_domain}

Considering the inflaton as a conformally coupled field, rapid-roll inflation is realized when three parameters ($\epsilon$, $\bar{\epsilon}$, $\eta_c$) are small, where $\epsilon$ is defined in Eq.~(\ref{SR-phi}), and $\bar{\epsilon}$ and $\eta_c$ in Eq.~(\ref{eqn:barepsilon-etac}). Defining $\delta \equiv (M_{\rm Pl}\Delta/\phi )^4$ and working with the Coulomb potential (\ref{eq_model_V}), i.e., setting $V(\phi)=V_C(\phi)$, the three parameters are
\begin{equation}
\epsilon = 8 \left(\frac{M_{\rm Pl}}{\phi}\right)^2 \left(\frac{\delta}{1-\delta}\right)^2,
\quad
\eta_c = -20 \left(\frac{M_{\rm Pl}}{\phi}\right)^2 \frac{\delta}{1-\delta},
\quad
\bar{\epsilon} = 2 \frac{\delta}{1-\delta}. 
\end{equation}
Here, since $V_{\phi\phi}\phi/V_{\phi}=-5$ is constant, we have set $c=5$ so that $\eta_c=\eta_{\phi}$. These three parameters are simultaneously small if $\phi / M_{\rm Pl}\gg \max[\Delta , \Delta^{2/3}]$, the first arising from $\epsilon_c \ll1$ and the second from $\vert \eta_\phi \vert \ll 1$. As geometrical reasons disfavor $\phi \gtrsim M_{\rm Pl}$ in D-brane inflation models (see, e.g., Refs.~\cite{Baumann:2006cd,Kobayashi:2007hm}), it is necessary to suppose that $\Delta \ll 1$. Assuming $\Delta \ll 1$, the condition for rapid-roll inflation is then $\vert \phi / M_{\rm Pl} \vert \gg \Delta^{2/3}$. The solution at the zeroth order in ($\epsilon$, $\bar{\epsilon}$, $\eta_c$) is given by Eq.~(\ref{eq_model_rr_phi}), and thus 
\begin{equation}
\partial_t{\phi} \simeq -H \phi.
\label{eq_rr_phi}
\end{equation}

The action (\ref{eq_model_rr_action}) is a first-order expansion in $(\partial_t{\phi})^2/(T_3h^4)$ of the DBI action with a conformal coupling to gravity, 
\begin{equation}
S_{DBI,cc} = 
\int \! \mathrm{d} ^{4} x \sqrt{-g}
\left[ \frac{1}{2} M_\mathrm{Pl}^{2} \mathcal{R} - T_{3} h^{4}(\phi) \sqrt{1 + \frac{\partial^{\mu} \phi \, \partial_{\mu} \phi}{T_{3}h^{4}(\phi)} }
+ T_{3} h^{4}(\phi) - V(\phi)
- \frac{\xi}{2} \mathcal{R} \phi^2 \right] \; ,
\label{eq_rr_DBI_cc_action}
\end{equation}
with $\xi = 1/6$ (see, e.g., Ref.~\cite{Easson:2009wc} for phenomenology of the nonminimally coupled DBI action). Thus, the action (\ref{eq_model_rr_action}) is justified only when $1-\gamma^{-2}\ll 1$, where $\gamma$ is defined by Eq.~(\ref{eqn:def_gamma}). Working with the $AdS_5$ throat $h = \phi/l$ and using $l^4 = \pi N T_3/[2 \mathrm{Vol}(X_5)]$, we have
\begin{equation}
 1- \frac{1}{\gamma^2} = \frac{(\partial_t\phi)^2}{T_3h^4} = \frac{(\partial_t\phi)^2}{\phi^4}\frac{\pi N}{2\mathrm{Vol}(X_5)}. 
\end{equation}
By substituting the rapid-roll behavior (\ref{eq_rr_phi}), we thus obtain
\begin{equation}
1 - \frac{1}{\gamma^2} = \frac{H^2}{\phi^2}\frac{\pi N}{2 \mathrm{Vol}(X_5)} = \sqrt\delta \left(\frac{H}{M_{\rm Pl}\Delta}\right)^2 \frac{\pi N}{2 \mathrm{Vol}(X_5)}.
\end{equation}
Hence, the behavior (\ref{eq_model_rr_phi}) implies that $1-\gamma^{-2}$ and thus $\gamma$ grows as the inflaton rapid-rolls.

For rapid-roll inflation, the authors of Ref.~\cite{Kofman:2007tr} gave the order of magnitude for the potential $V_0 \sim (\mathrm{TeV})^4$, so $H/M_{\rm Pl} \sim (\mathrm{TeV}/M_{\rm Pl})^2 \sim 10^{-30}$. Assuming $\pi N/[2 \mathrm{Vol}(X_5)] = \mathcal{O}(1)$ leads to $1 - \gamma^{-2} \sim 10^{-60} \sqrt\delta/\Delta^2 \sim 10^{-60} (M_{\rm Pl}/\phi)^2 \ll 10^{-60}\Delta^{-4/3}$, where the condition $\phi / M_{\rm Pl}\gg \Delta^{2/3}$ was used. Thus, although $\gamma$ grows, $1-\gamma^{-2}$ remains small all the way down to the end of the rapid-roll inflation, provided that $\Delta \gtrsim 10^{-23}$, which is compatible with the assumption $\Delta \ll 1$. Hereafter, it will thus be assumed that
\begin{equation}
10^{-23} \lesssim \Delta \ll 1.
\end{equation}

\subsection{Ruling out a brane curvaton model}
\label{subsec:rapidroll-branecurvaton}

The behavior of the brane curvaton in rapid-roll inflation is quite peculiar. As we shall show below, the curvaton on a brane produces a strongly blue spectrum of the curvature perturbations when the inflation is driven by a rapidly rolling inflaton and thus is disfavored by the observational data.

At the zeroth order in $\delta$, the scale factor of the brane metric is $\tilde{a} = h_0 a_0 (a/a_0)^\beta$. Thus, for the $AdS_5$ throat ($\beta = 0$), the brane metric does not expand. We thus need to take into account the effects of $\delta\ne 0$ and/or $\beta\ne 0$. Let us therefore go to the next order in $\delta$. First, combining the equations in Eq.~(\ref{eq_model_rr_EOM}) yields the master equation for $\phi$,	
\begin{equation}
\partial_t \phi  = - \left(1-\frac{3\eta_\phi}{35}\right) H\phi.
\label{eq_rr_master}
\end{equation}
The solution of this equation at the first order in $\delta$ is
\begin{equation}
\phi(t) \simeq \phi_0 \frac{a_0}{a} - \frac{2}{7} \, \frac{M_{\rm Pl}^2 (M_{\rm Pl} \Delta)^4}{\phi_0^5} \left( \frac{a}{a_0} \right)^5 \simeq
\phi_0 \frac{a_0}{a} \left[ 1 - \frac{2\delta}{7} \, \frac{M_{\rm Pl}^2}{\phi_0^2} \left( \frac{a}{a_0} \right)^2 \right] \; ,
\end{equation}
which satisfies Eq.~(\ref{eq_model_rr_EOM}) with $c=5$. The brane frame metric at this order is then 
\begin{equation}
\tilde{N} = \frac{h_0a_0}{\gamma a} \left(\frac{a}{a_{0}}\right)^{\beta} ,
\quad
\tilde{a} = h_0a_0\left(\frac{a}{a_{0}}\right)^{\beta} 
\left[ 1 - \frac{2\delta}{7} \, \frac{M_{\rm Pl}^2}{\phi_0^2} \left( \frac{a}{a_0} \right)^2 \right]^{1-\beta} \; .
\label{eq_rr_brane_qtt}
\end{equation}
Unfortunately, for $\beta=0$ and $\delta > 0$, the brane metric does not expand, but rather contracts.

Thus, we next take into account $\beta\ne 0$, i.e., a deviation from the $AdS_5$ throat. For simplicity, we shall neglect the $\delta$ and $\gamma^2 \! - \! 1$ contributions. The brane metric is now given by 
\begin{equation}
\tilde{a} \simeq h_0a_0\left(\frac{a}{a_{0}}\right)^{\beta},
\quad
\tilde{N} \simeq \frac{\tilde{a}}{a},
\quad
\tilde{H} \simeq \frac{\beta H}{\tilde{N}},
\quad
\tilde{\epsilon} \simeq \frac{\beta - 1}{\beta}.
\end{equation}
Thus, the brane metric expands if $\beta>0$. The EOM in Eq.~(\ref{eq_model_EOM_t}) for a Fourier mode of the curvaton perturbation, $\delta \hat{\sigma}_\mathbf{k} = w_k \hat{a}_\mathbf{k} \! + \! w_k^* \hat{a}^\dagger_\mathbf{k}$, is 
\begin{equation}
 \partial_{\tilde{\eta}}^2 w_k + \frac{2\beta}{-\tilde{\eta}}\partial_{\tilde{\eta}} w_k +K^2 w_k = 0, \label{eq_rr_EOM_wk}
\end{equation}
where $K^{2} \equiv k^{2} +  \tilde{a}^{2} V_{\sigma \sigma}$ and $\tilde\eta$ is the conformal time on the brane, defined in Eq.~\eqref{def-etatil}. Taking the massless limit for simplicity, $K^2=k^2$, we obtain the Bunch-Davies type solution in terms of the Hankel function as
\begin{equation}
w_k = \frac{i}{\tilde a} \, \frac{\sqrt{\pi}}{2} \sqrt{- \tilde{\eta}} \, H^{(1)}_{\beta+1/2} ( - k \tilde{\eta}) \; ,
\end{equation}
up to an irrelevant constant phase, where $H^{(1)}_\nu (x)$ is the Hankel function of the first kind. In the superhorizon limit, $-k \tilde{\eta} \ll 1$, this expression reduces to
\begin{equation}
w_{k} \simeq \frac{1}{a_{0}h_{0}} \frac{(2 H_{\tilde *} a_{0})^{\beta}}{\sqrt{2\pi}} \frac{\Gamma
(\frac{1}{2} \! + \!\beta)}{k^{\frac{1}{2} + \beta}}.
\end{equation}
Following Ref.~\cite{Lyth:2001nq}, this mode leads to the power spectrum and the spectral index, 
\begin{equation}
\tilde{\mathcal{P}}_{\zeta} = \frac{(2 H_{\tilde *} a_{0})^{2\beta}}{\pi ( a_{0} h_{0})^{2}} \left( \frac{R \Gamma(\frac{1}{2} \! + \!\beta) }{3 \pi \sigma_*}\right)^{2} k^{\tilde{n}_s - 1}, 
\quad
\tilde{n}_s - 1 =  2(1 - \beta).
\label{eq_rr_brane_spectrum}
\end{equation}
To derive Eq.~\eqref{eq_rr_brane_spectrum}, $H$ has been considered as a constant. In more general situations, an extra $k$ dependence is expected from $H_{\tilde{*}}$.

With a small deviation from the $AdS_5 \times X^5$ geometry (i.e., small $\beta$), the brane curvaton model in rapid-roll inflation induces a too large scale dependence, $\tilde{n}_s \simeq 3$, while current data constrain $\tilde{n}_s = 0.968 \pm 0.006$ \cite{Ade:2015lrj}.

\subsection{Bulk curvaton model}\label{sec_rr_bulk}

In the bulk curvaton model, assuming the mass hierarchy $m_\sigma \ll H$ (i.e., $\eta_{\sigma} \ll 1$), the power spectrum and the spectral index are given by Eq.~(\ref{eq_model_Pc}). One thing that we should notice is that, unlike slow-roll inflation, we have $\epsilon_H\simeq \bar{\epsilon}$~\cite{Kofman:2007tr}. Thus, the spectral index in the bulk curvaton model is given by 
\begin{equation}
 n_s-1 \simeq 2\eta_{\sigma}- 2 \bar{\epsilon}.
\end{equation}
Thus the observational constraint $\tilde{n}_s = 0.968 \pm 0.006$ \cite{Ade:2015lrj} can be fulfilled in this model.

\section{Summary and discussion}
\label{sec:summary}

The invariance of primordial curvature perturbations under conformal/disformal transformations has been known independent of which metric/frame the standard model particles couple to, provided that the perturbations are originated from the inflaton fluctuations \cite{Deruelle:2010ht,Gong:2011qe,Chiba:2013mha,Motohashi:2015pra,Domenech:2015hka}. In this work we have considered the effects of another field being responsible for the production of curvature perturbations---dubbed the curvaton---comparing two different metrics/frames to which it minimally couples to. We have explicitly demonstrated in concrete setups of D-brane inflation that this difference indeed leaves a gap in the observables, specifically the scalar spectral index, showing the falsifiability or possibly alleviating the difficulty in the original models without the curvaton.

In the context of a slow-roll D-brane inflation with an inflection-point potential and a rapid-roll D-brane inflation with a conformal coupling, we have considered the difference between the two frames/metrics in the curvaton scenario and have studied cosmological perturbations originated from each of them. In the bulk curvaton model the curvaton minimally couples to the bulk metric $g_{\mu\nu}$, which is the four-dimensional part of the ten-dimensional bulk metric, as in Eq.~(\ref{eq_model_ds_10}). In the brane curvaton model, on the other hand, the curvaton minimally couples to the brane metric $\tilde{g}_{\mu\nu}$, which is the four-dimensional induced metric on the world volume of the inflaton D-brane, as in Eq.~(\ref{eq_model_gmnt}). In the case of slow-roll inflation with an inflection-point potential, we have shown that the difference of the spectral indices between the two curvaton models is not necessarily small, as shown in Eq.~(\ref{eqn:ns-nstilde-inflection-AdS5}). Thus, in this case the inflaton potential, the curvaton mass, and the warp factor are constrained by observation in different ways in the two curvaton models. Moreover, with a curvaton, the amount of fine-tuning of the inflaton potential is significantly relaxed.  In the case of conformal rapid-roll inflation, while the bulk curvaton model can be consistent with observational data, we have shown in Sec.~\ref{subsec:rapidroll-branecurvaton} that the brane curvaton model is excluded by observation. The two metrics $g_{\mu\nu}$ and $\tilde g_{\mu\nu}$ are related by a conformal/disformal transformation, and thus our present study provides an explicit example for which such a metric transformation has a physical impact on observable quantities.

The two models of inflation considered in the present paper are low-energy models and thus predict a small tensor-to-scalar ratio well below the current bound $r < 0.11$ (95\% C.L.) \cite{Ade:2015lrj}. The introduction of a curvaton as the dominant source of curvature perturbations does not change this conclusion and actually reduces the tensor-to-scalar ratio to an even smaller value. On the other hand, the curvaton scenario may yield a relatively large non-Gaussianity~\cite{Sasaki:2006kq}, unlike the standard single-field inflation models~\cite{Maldacena:2002vr}. In the absence of nonlinear evolution and for $R=1$ (see the definition of $R$ after Eq.~(\ref{Pzeta_gen})), the nonlinear parameter is predicted to be $f_{NL}=5/4$. The current bound $f_{NL} = 0.8 \pm 5.0$ (68\% C.L.) \cite{Ade:2015ava} indicates that curvaton models may observationally be either excluded or confirmed in the near future.

\begin{acknowledgments}
F.\,L.\, is grateful to Yukawa Institute for Theoretical Physics (YITP) for hosting him during his internship. This work was supported in part by Japan Society for the Promotion of Science (JSPS) Grants-in-Aid for Scientific Research No.\,24540256 (S.\,M.) and No.\,16J06266 (Y.\,W.), by the Program for Leading Graduate Schools, MEXT, Japan (Y.\,W.), and by World Premier International Research Center Initiative (WPI), MEXT, Japan. 
\end{acknowledgments}


\begin{thebibliography}{99}

%\cite{Guth:1980zm}
\bibitem{Guth:1980zm} 
  A.~H.~Guth,
  ``The Inflationary Universe: A Possible Solution to the Horizon and Flatness Problems,''
  Phys.\ Rev.\ D {\bf 23}, 347 (1981).
  doi:10.1103/PhysRevD.23.347
  %%CITATION = doi:10.1103/PhysRevD.23.347;%%
  %6117 citations counted in INSPIRE as of 14 Jan 2017


%\cite{Sato:1980yn}
\bibitem{Sato:1980yn} 
  K.~Sato,
  ``First Order Phase Transition of a Vacuum and Expansion of the Universe,''
  Mon.\ Not.\ Roy.\ Astron.\ Soc.\  {\bf 195}, 467 (1981).
  %%CITATION = MNRAA,195,467;%%
  %965 citations counted in INSPIRE as of 14 Jan 2017


%\cite{Starobinsky:1980te}
\bibitem{Starobinsky:1980te} 
  A.~A.~Starobinsky,
  ``A New Type of Isotropic Cosmological Models Without Singularity,''
  Phys.\ Lett.\  {\bf 91B}, 99 (1980).
  doi:10.1016/0370-2693(80)90670-X
  %%CITATION = doi:10.1016/0370-2693(80)90670-X;%%
  %3034 citations counted in INSPIRE as of 14 Jan 2017


%\cite{Inflation_String_th}
\bibitem{Inflation_String_th}
D.~Baumann and L.~McAllister, "Inflation and String Theory", Cambridge University Press, 2015.


%\cite{Dvali:1998pa}
\bibitem{Dvali:1998pa} 
  G.~R.~Dvali and S.~H.~H.~Tye,
  ``Brane inflation,''
  Phys.\ Lett.\ B {\bf 450}, 72 (1999)
  doi:10.1016/S0370-2693(99)00132-X
  [hep-ph/9812483].
  %%CITATION = doi:10.1016/S0370-2693(99)00132-X;%%
  %794 citations counted in INSPIRE as of 14 Jan 2017


%\cite{Dvali:2001fw}
\bibitem{Dvali:2001fw} 
  G.~R.~Dvali, Q.~Shafi and S.~Solganik,
  ``D-brane inflation,''
  hep-th/0105203.
  %%CITATION = HEP-TH/0105203;%%
  %292 citations counted in INSPIRE as of 14 Jan 2017


%\cite{Burgess:2001fx}
\bibitem{Burgess:2001fx} 
  C.~P.~Burgess, M.~Majumdar, D.~Nolte, F.~Quevedo, G.~Rajesh and R.~J.~Zhang,
  ``The Inflationary brane anti-brane universe,''
  JHEP {\bf 0107}, 047 (2001)
  doi:10.1088/1126-6708/2001/07/047
  [hep-th/0105204].
  %%CITATION = doi:10.1088/1126-6708/2001/07/047;%%
  %446 citations counted in INSPIRE as of 14 Jan 2017


%\cite{Kobayashi:2007hm}
\bibitem{Kobayashi:2007hm} 
  T.~Kobayashi, S.~Mukohyama and S.~Kinoshita,
  ``Constraints on Wrapped DBI Inflation in a Warped Throat,''
  JCAP {\bf 0801}, 028 (2008)
  doi:10.1088/1475-7516/2008/01/028
  [arXiv:0708.4285 [hep-th]].
  %%CITATION = doi:10.1088/1475-7516/2008/01/028;%%
  %59 citations counted in INSPIRE as of 14 Jan 2017


%\cite{Kachru:2003sx}
\bibitem{Kachru:2003sx} 
  S.~Kachru, R.~Kallosh, A.~D.~Linde, J.~M.~Maldacena, L.~P.~McAllister and S.~P.~Trivedi,
  ``Towards inflation in string theory,''
  JCAP {\bf 0310}, 013 (2003)
  doi:10.1088/1475-7516/2003/10/013
  [hep-th/0308055].
  %%CITATION = doi:10.1088/1475-7516/2003/10/013;%%
  %1004 citations counted in INSPIRE as of 14 Jan 2017


%\cite{Baumann:2007np}
\bibitem{Baumann:2007np} 
  D.~Baumann, A.~Dymarsky, I.~R.~Klebanov, L.~McAllister and P.~J.~Steinhardt,
  ``A Delicate universe,''
  Phys.\ Rev.\ Lett.\  {\bf 99}, 141601 (2007)
  doi:10.1103/PhysRevLett.99.141601
  [arXiv:0705.3837 [hep-th]].
  %%CITATION = doi:10.1103/PhysRevLett.99.141601;%%
  %151 citations counted in INSPIRE as of 14 Jan 2017


%\cite{Kofman:2007tr}
\bibitem{Kofman:2007tr} 
  L.~Kofman and S.~Mukohyama,
  ``Rapid roll Inflation with Conformal Coupling,''
  Phys.\ Rev.\ D {\bf 77}, 043519 (2008)
  doi:10.1103/PhysRevD.77.043519
  [arXiv:0709.1952 [hep-th]].
  %%CITATION = doi:10.1103/PhysRevD.77.043519;%%
  %36 citations counted in INSPIRE as of 14 Jan 2017


%\cite{Enqvist:2001zp}
\bibitem{Enqvist:2001zp} 
  K.~Enqvist and M.~S.~Sloth,
  ``Adiabatic CMB perturbations in pre - big bang string cosmology,''
  Nucl.\ Phys.\ B {\bf 626}, 395 (2002)
  doi:10.1016/S0550-3213(02)00043-3
  [hep-ph/0109214].
  %%CITATION = doi:10.1016/S0550-3213(02)00043-3;%%
  %652 citations counted in INSPIRE as of 14 Jan 2017


%\cite{Lyth:2001nq}
\bibitem{Lyth:2001nq} 
  D.~H.~Lyth and D.~Wands,
  ``Generating the curvature perturbation without an inflaton,''
  Phys.\ Lett.\ B {\bf 524}, 5 (2002)
  doi:10.1016/S0370-2693(01)01366-1
  [hep-ph/0110002].
  %%CITATION = doi:10.1016/S0370-2693(01)01366-1;%%
  %1048 citations counted in INSPIRE as of 14 Jan 2017


%\cite{Moroi:2001ct}
\bibitem{Moroi:2001ct} 
  T.~Moroi and T.~Takahashi,
  ``Effects of cosmological moduli fields on cosmic microwave background,''
  Phys.\ Lett.\ B {\bf 522}, 215 (2001)
  Erratum: [Phys.\ Lett.\ B {\bf 539}, 303 (2002)]
  doi:10.1016/S0370-2693(02)02070-1, 10.1016/S0370-2693(01)01295-3
  [hep-ph/0110096].
  %%CITATION = doi:10.1016/S0370-2693(02)02070-1, 10.1016/S0370-2693(01)01295-3;%%
  %753 citations counted in INSPIRE as of 14 Jan 2017


%\cite{Bekenstein:1992pj}
\bibitem{Bekenstein:1992pj} 
  J.~D.~Bekenstein,
  ``The Relation between physical and gravitational geometry,''
  Phys.\ Rev.\ D {\bf 48}, 3641 (1993)
  doi:10.1103/PhysRevD.48.3641
  [gr-qc/9211017].
  %%CITATION = doi:10.1103/PhysRevD.48.3641;%%
  %178 citations counted in INSPIRE as of 14 Jan 2017


%\cite{Deruelle:2010ht}
\bibitem{Deruelle:2010ht} 
  N.~Deruelle and M.~Sasaki,
  ``Conformal equivalence in classical gravity: the example of 'Veiled' General Relativity,''
  Springer Proc.\ Phys.\  {\bf 137}, 247 (2011)
  doi:10.1007/978-3-642-19760-4\_23
  [arXiv:1007.3563 [gr-qc]].
  %%CITATION = doi:10.1007/978-3-642-19760-4_23;%%
  %66 citations counted in INSPIRE as of 14 Jan 2017


%\cite{Gong:2011qe}
\bibitem{Gong:2011qe} 
  J.~O.~Gong, J.~c.~Hwang, W.~I.~Park, M.~Sasaki and Y.~S.~Song,
  ``Conformal invariance of curvature perturbation,''
  JCAP {\bf 1109}, 023 (2011)
  doi:10.1088/1475-7516/2011/09/023
  [arXiv:1107.1840 [gr-qc]].
  %%CITATION = doi:10.1088/1475-7516/2011/09/023;%%
  %43 citations counted in INSPIRE as of 14 Jan 2017


%\cite{Chiba:2013mha}
\bibitem{Chiba:2013mha} 
  T.~Chiba and M.~Yamaguchi,
  ``Conformal-Frame (In)dependence of Cosmological Observations in Scalar-Tensor Theory,''
  JCAP {\bf 1310}, 040 (2013)
  doi:10.1088/1475-7516/2013/10/040
  [arXiv:1308.1142 [gr-qc]].
  %%CITATION = doi:10.1088/1475-7516/2013/10/040;%%
  %40 citations counted in INSPIRE as of 14 Jan 2017


%\cite{Motohashi:2015pra}
\bibitem{Motohashi:2015pra} 
  H.~Motohashi and J.~White,
  ``Disformal invariance of curvature perturbation,''
  JCAP {\bf 1602}, no. 02, 065 (2016)
  doi:10.1088/1475-7516/2016/02/065
  [arXiv:1504.00846 [gr-qc]].
  %%CITATION = doi:10.1088/1475-7516/2016/02/065;%%
  %15 citations counted in INSPIRE as of 14 Jan 2017


%\cite{Domenech:2015hka}
\bibitem{Domenech:2015hka} 
  G.~Dom\`enech, A.~Naruko and M.~Sasaki,
  ``Cosmological disformal invariance,''
  JCAP {\bf 1510}, no. 10, 067 (2015)
  doi:10.1088/1475-7516/2015/10/067
  [arXiv:1505.00174 [gr-qc]].
  %%CITATION = doi:10.1088/1475-7516/2015/10/067;%%
  %30 citations counted in INSPIRE as of 14 Jan 2017


%\cite{Domenech:2015tca}
\bibitem{Domenech:2015tca} 
  G.~Dom\`enech, S.~Mukohyama, R.~Namba, A.~Naruko, R.~Saitou and Y.~Watanabe,
  ``Derivative-dependent metric transformation and physical degrees of freedom,''
  Phys.\ Rev.\ D {\bf 92}, no. 8, 084027 (2015)
  doi:10.1103/PhysRevD.92.084027
  [arXiv:1507.05390 [hep-th]].
  %%CITATION = doi:10.1103/PhysRevD.92.084027;%%
  %27 citations counted in INSPIRE as of 14 Jan 2017


%\cite{Klebanov:2000hb}
\bibitem{Klebanov:2000hb} 
  I.~R.~Klebanov and M.~J.~Strassler,
  ``Supergravity and a confining gauge theory: Duality cascades and $\chi$SB-resolution of naked singularities,''
  JHEP {\bf 0008}, 052 (2000)
  doi:10.1088/1126-6708/2000/08/052
  [hep-th/0007191].
  %%CITATION = doi:10.1088/1126-6708/2000/08/052;%%
  %1502 citations counted in INSPIRE as of 14 Jan 2017


%\cite{Kuperstein:2004hy}
\bibitem{Kuperstein:2004hy} 
  S.~Kuperstein,
  ``Meson spectroscopy from holomorphic probes on the warped deformed conifold,''
  JHEP {\bf 0503}, 014 (2005)
  doi:10.1088/1126-6708/2005/03/014
  [hep-th/0411097].
  %%CITATION = doi:10.1088/1126-6708/2005/03/014;%%
  %128 citations counted in INSPIRE as of 14 Jan 2017


%\cite{Ade:2015lrj}
\bibitem{Ade:2015lrj} 
  P.~A.~R.~Ade {\it et al.} [Planck Collaboration],
  ``Planck 2015 results. XX. Constraints on inflation,''
  Astron.\ Astrophys.\  {\bf 594}, A20 (2016)
  doi:10.1051/0004-6361/201525898
  [arXiv:1502.02114 [astro-ph.CO]].
  %%CITATION = doi:10.1051/0004-6361/201525898;%%
  %952 citations counted in INSPIRE as of 14 Jan 2017


%\cite{Giddings:2001yu}
\bibitem{Giddings:2001yu} 
  S.~B.~Giddings, S.~Kachru and J.~Polchinski,
  ``Hierarchies from fluxes in string compactifications,''
  Phys.\ Rev.\ D {\bf 66}, 106006 (2002)
  doi:10.1103/PhysRevD.66.106006
  [hep-th/0105097].
  %%CITATION = doi:10.1103/PhysRevD.66.106006;%%
  %1607 citations counted in INSPIRE as of 14 Jan 2017


%\cite{Baumann:2007ah}
\bibitem{Baumann:2007ah} 
  D.~Baumann, A.~Dymarsky, I.~R.~Klebanov and L.~McAllister,
  ``Towards an Explicit Model of D-brane Inflation,''
  JCAP {\bf 0801}, 024 (2008)
  doi:10.1088/1475-7516/2008/01/024
  [arXiv:0706.0360 [hep-th]].
  %%CITATION = doi:10.1088/1475-7516/2008/01/024;%%
  %200 citations counted in INSPIRE as of 14 Jan 2017


%\cite{Agarwal:2011wm}
\bibitem{Agarwal:2011wm} 
  N.~Agarwal, R.~Bean, L.~McAllister and G.~Xu,
  ``Universality in D-brane Inflation,''
  JCAP {\bf 1109}, 002 (2011)
  doi:10.1088/1475-7516/2011/09/002
  [arXiv:1103.2775 [astro-ph.CO]].
  %%CITATION = doi:10.1088/1475-7516/2011/09/002;%%
  %63 citations counted in INSPIRE as of 14 Jan 2017


%\cite{Dias:2012nf}
\bibitem{Dias:2012nf} 
  M.~Dias, J.~Frazer and A.~R.~Liddle,
  ``Multifield consequences for D-brane inflation,''
  JCAP {\bf 1206}, 020 (2012)
  Erratum: [JCAP {\bf 1303}, E01 (2013)]
  doi:10.1088/1475-7516/2013/03/E01, 10.1088/1475-7516/2012/06/020
  [arXiv:1203.3792 [astro-ph.CO]].
  %%CITATION = doi:10.1088/1475-7516/2013/03/E01, 10.1088/1475-7516/2012/06/020;%%
  %33 citations counted in INSPIRE as of 14 Jan 2017


%\cite{McAllister:2012am}
\bibitem{McAllister:2012am} 
  L.~McAllister, S.~Renaux-Petel and G.~Xu,
  ``A Statistical Approach to Multifield Inflation: Many-field Perturbations Beyond Slow Roll,''
  JCAP {\bf 1210}, 046 (2012)
  doi:10.1088/1475-7516/2012/10/046
  [arXiv:1207.0317 [astro-ph.CO]].
  %%CITATION = doi:10.1088/1475-7516/2012/10/046;%%
  %52 citations counted in INSPIRE as of 14 Jan 2017


%\cite{Baumann:2010sx}
\bibitem{Baumann:2010sx} 
  D.~Baumann, A.~Dymarsky, S.~Kachru, I.~R.~Klebanov and L.~McAllister,
  ``D3-brane Potentials from Fluxes in AdS/CFT,''
  JHEP {\bf 1006}, 072 (2010)
  doi:10.1007/JHEP06(2010)072
  [arXiv:1001.5028 [hep-th]].
  %%CITATION = doi:10.1007/JHEP06(2010)072;%%
  %95 citations counted in INSPIRE as of 14 Jan 2017


%\cite{Gandhi:2011id}
\bibitem{Gandhi:2011id} 
  S.~Gandhi, L.~McAllister and S.~Sjors,
  ``A Toolkit for Perturbing Flux Compactifications,''
  JHEP {\bf 1112}, 053 (2011)
  doi:10.1007/JHEP12(2011)053
  [arXiv:1106.0002 [hep-th]].
  %%CITATION = doi:10.1007/JHEP12(2011)053;%%
  %9 citations counted in INSPIRE as of 14 Jan 2017


%\cite{Panda:2007ie}
\bibitem{Panda:2007ie} 
  S.~Panda, M.~Sami and S.~Tsujikawa,
  ``Prospects of inflation in delicate D-brane cosmology,''
  Phys.\ Rev.\ D {\bf 76}, 103512 (2007)
  doi:10.1103/PhysRevD.76.103512
  [arXiv:0707.2848 [hep-th]].
  %%CITATION = doi:10.1103/PhysRevD.76.103512;%%
  %51 citations counted in INSPIRE as of 14 Jan 2017


%\cite{Krause:2007jk}
\bibitem{Krause:2007jk} 
  A.~Krause and E.~Pajer,
  ``Chasing brane inflation in string-theory,''
  JCAP {\bf 0807}, 023 (2008)
  doi:10.1088/1475-7516/2008/07/023
  [arXiv:0705.4682 [hep-th]].
  %%CITATION = doi:10.1088/1475-7516/2008/07/023;%%
  %100 citations counted in INSPIRE as of 14 Jan 2017


%\cite{Shandera:2006ax}
\bibitem{Shandera:2006ax} 
  S.~E.~Shandera and S.-H.~H.~Tye,
  ``Observing brane inflation,''
  JCAP {\bf 0605}, 007 (2006)
  doi:10.1088/1475-7516/2006/05/007
  [hep-th/0601099].
  %%CITATION = doi:10.1088/1475-7516/2006/05/007;%%
  %112 citations counted in INSPIRE as of 14 Jan 2017


%\cite{Peiris:2007gz}
\bibitem{Peiris:2007gz} 
  H.~V.~Peiris, D.~Baumann, B.~Friedman and A.~Cooray,
  ``Phenomenology of D-Brane Inflation with General Speed of Sound,''
  Phys.\ Rev.\ D {\bf 76}, 103517 (2007)
  doi:10.1103/PhysRevD.76.103517
  [arXiv:0706.1240 [astro-ph]].
  %%CITATION = doi:10.1103/PhysRevD.76.103517;%%
  %88 citations counted in INSPIRE as of 14 Jan 2017


%\cite{Chen:2008ai}
\bibitem{Chen:2008ai} 
  H.~Y.~Chen and J.~O.~Gong,
  ``Towards a warped inflationary brane scanning,''
  Phys.\ Rev.\ D {\bf 80}, 063507 (2009)
  doi:10.1103/PhysRevD.80.063507
  [arXiv:0812.4649 [hep-th]].
  %%CITATION = doi:10.1103/PhysRevD.80.063507;%%
  %12 citations counted in INSPIRE as of 14 Jan 2017


%\cite{Chen:2008ada}
\bibitem{Chen:2008ada} 
  H.~Y.~Chen, J.~O.~Gong and G.~Shiu,
  ``Systematics of multi-field effects at the end of warped brane inflation,''
  JHEP {\bf 0809}, 011 (2008)
  doi:10.1088/1126-6708/2008/09/011
  [arXiv:0807.1927 [hep-th]].
  %%CITATION = doi:10.1088/1126-6708/2008/09/011;%%
  %28 citations counted in INSPIRE as of 14 Jan 2017


%\cite{Pilo:2004mg}
\bibitem{Pilo:2004mg} 
  L.~Pilo, A.~Riotto and A.~Zaffaroni,
  ``Old inflation in string theory,''
  JHEP {\bf 0407}, 052 (2004)
  doi:10.1088/1126-6708/2004/07/052
  [hep-th/0401004].
  %%CITATION = doi:10.1088/1126-6708/2004/07/052;%%
  %42 citations counted in INSPIRE as of 14 Jan 2017


%\cite{Li:2008fma}
\bibitem{Li:2008fma} 
  Y.~F.~Cai, S.~Li and Y.~S.~Piao,
  ``DBI-Curvaton,''
  Phys.\ Lett.\ B {\bf 671}, 423 (2009)
  doi:10.1016/j.physletb.2008.12.056
  [arXiv:0806.2363 [hep-ph]].
  %%CITATION = doi:10.1016/j.physletb.2008.12.056;%%
  %51 citations counted in INSPIRE as of 14 Jan 2017


%\cite{Cai:2010rt}
\bibitem{Cai:2010rt} 
  Y.~F.~Cai and Y.~Wang,
  ``Large Nonlocal Non-Gaussianity from a Curvaton Brane,''
  Phys.\ Rev.\ D {\bf 82}, 123501 (2010)
  doi:10.1103/PhysRevD.82.123501
  [arXiv:1005.0127 [hep-th]].
  %%CITATION = doi:10.1103/PhysRevD.82.123501;%%
  %24 citations counted in INSPIRE as of 14 Jan 2017


%\cite{Kobayashi:2009cm}
\bibitem{Kobayashi:2009cm} 
  T.~Kobayashi and S.~Mukohyama,
  ``Curvatons in Warped Throats,''
  JCAP {\bf 0907}, 032 (2009)
  doi:10.1088/1475-7516/2009/07/032
  [arXiv:0905.2835 [hep-th]].
  %%CITATION = doi:10.1088/1475-7516/2009/07/032;%%
  %16 citations counted in INSPIRE as of 14 Jan 2017


%\cite{Baumann:2006cd}
\bibitem{Baumann:2006cd} 
  D.~Baumann and L.~McAllister,
  ``A Microscopic Limit on Gravitational Waves from D-brane Inflation,''
  Phys.\ Rev.\ D {\bf 75}, 123508 (2007)
  doi:10.1103/PhysRevD.75.123508
  [hep-th/0610285].
  %%CITATION = doi:10.1103/PhysRevD.75.123508;%%
  %178 citations counted in INSPIRE as of 14 Jan 2017


%\cite{Easson:2009wc}
\bibitem{Easson:2009wc} 
  D.~A.~Easson, S.~Mukohyama and B.~A.~Powell,
  ``Observational Signatures of Gravitational Couplings in DBI Inflation,''
  Phys.\ Rev.\ D {\bf 81}, 023512 (2010)
  doi:10.1103/PhysRevD.81.023512
  [arXiv:0910.1353 [astro-ph.CO]].
  %%CITATION = doi:10.1103/PhysRevD.81.023512;%%
  %25 citations counted in INSPIRE as of 14 Jan 2017


%\cite{Sasaki:2006kq}
\bibitem{Sasaki:2006kq} 
  M.~Sasaki, J.~Valiviita and D.~Wands,
  ``Non-Gaussianity of the primordial perturbation in the curvaton model,''
  Phys.\ Rev.\ D {\bf 74}, 103003 (2006)
  doi:10.1103/PhysRevD.74.103003
  [astro-ph/0607627].
  %%CITATION = doi:10.1103/PhysRevD.74.103003;%%
  %281 citations counted in INSPIRE as of 14 Jan 2017


%\cite{Maldacena:2002vr}
\bibitem{Maldacena:2002vr} 
  J.~M.~Maldacena,
  ``Non-Gaussian features of primordial fluctuations in single field inflationary models,''
  JHEP {\bf 0305}, 013 (2003)
  doi:10.1088/1126-6708/2003/05/013
  [astro-ph/0210603].
  %%CITATION = doi:10.1088/1126-6708/2003/05/013;%%
  %1590 citations counted in INSPIRE as of 14 Jan 2017


%\cite{Ade:2015ava}
\bibitem{Ade:2015ava} 
  P.~A.~R.~Ade {\it et al.} [Planck Collaboration],
  ``Planck 2015 results. XVII. Constraints on primordial non-Gaussianity,''
  Astron.\ Astrophys.\  {\bf 594}, A17 (2016)
  doi:10.1051/0004-6361/201525836
  [arXiv:1502.01592 [astro-ph.CO]].
  %%CITATION = doi:10.1051/0004-6361/201525836;%%
  %289 citations counted in INSPIRE as of 14 Jan 2017


\end{thebibliography}
\end{document}